\documentclass[10pt,conference,letterpaper,nofonttune]{IEEEtran}  %a4paper
\IEEEoverridecommandlockouts
\usepackage{amsmath,amssymb,amsfonts}
\usepackage[ruled,vlined]{algorithm2e}
\usepackage{graphicx}
\usepackage{textcomp}
\usepackage{xcolor}
\usepackage{soul}
\usepackage{balance}

\usepackage{glossaries}
\newacronym{3gpp}{3GPP}{3rd Generation Partnership Project}
\newacronym{4g}{4G}{4th generation mobile network}
\newacronym{5g}{5G}{5th generation mobile network}
\newacronym{6g}{6G}{6th generation mobile network}
\newacronym{nextg}{NextG}{Next Generation}
\newacronym{5gc}{5GC}{5G Core}
\newacronym{adc}{ADC}{Analog to Digital Converter}
\newacronym{aerpaw}{AERPAW}{Aerial Experimentation and Research Platform for Advanced Wireless}
\newacronym{ai}{AI}{Artificial Intelligence}
\newacronym{aimd}{AIMD}{Additive Increase Multiplicative Decrease}
\newacronym{am}{AM}{Acknowledged Mode}
\newacronym{amc}{AMC}{Adaptive Modulation and Coding}
\newacronym{amf}{AMF}{Access and Mobility Management Function}
\newacronym{aops}{AOPS}{Adaptive Order Prediction Scheduling}
\newacronym{api}{API}{Application Programming Interface}
\newacronym{apn}{APN}{Access Point Name}
\newacronym{aqm}{AQM}{Active Queue Management}
\newacronym{ausf}{AUSF}{Authentication Server Function}
\newacronym{avc}{AVC}{Advanced Video Coding}
\newacronym{awgn}{AGWN}{Additive White Gaussian Noise}
\newacronym{balia}{BALIA}{Balanced Link Adaptation Algorithm}
\newacronym{bbu}{BBU}{Base Band Unit}
\newacronym{bdp}{BDP}{Bandwidth-Delay Product}
\newacronym{ber}{BER}{Bit Error Rate}
\newacronym{bf}{BF}{Beamforming}
\newacronym{bler}{BLER}{Block Error Rate}
\newacronym{brr}{BRR}{Bayesian Ridge Regressor}
\newacronym{bsr}{BSR}{Buffer Status Report}
\newacronym{bs}{BS}{Base Station}
\newacronym{bpsk}{BPSK}{Binary Phase-shift keying}
\newacronym{bss}{BSS}{Business Support System}
\newacronym{ca}{CA}{Carrier Aggregation}
\newacronym{caas}{CaaS}{Connectivity-as-a-Service}
\newacronym{cb}{CB}{Code Block}
\newacronym{cc}{CC}{Congestion Control}
\newacronym{ccid}{CCID}{Congestion Control ID}
\newacronym{cco}{CC}{Carrier Component}
\newacronym{cd}{CD}{Continuous Delivery}
\newacronym{cdd}{CDD}{Cyclic Delay Diversity}
\newacronym{cdf}{CDF}{Cumulative Distribution Function}
\newacronym{cdma}{CDMA}{Code-Division Multiple Access}
\newacronym{cdn}{CDN}{Content Distribution Network}
\newacronym{ci}{CI}{Continuous Integration}
\newacronym{cicd}{CI/CD}{Continuous Integration/Continuous Delivery}
\newacronym{cir}{CIR}{Channel Impulse Response}
\newacronym{cn}{CN}{Core Network}
\newacronym{codel}{CoDel}{Controlled Delay Management}
\newacronym{comac}{COMAC}{Converged Multi-Access and Core}
\newacronym{cord}{CORD}{Central Office Re-architected as a Datacenter}
\newacronym{cornet}{CORNET}{COgnitive Radio NETwork}
\newacronym{cosmos}{COSMOS}{Cloud Enhanced Open Software Defined Mobile Wireless Testbed for City-Scale Deployment}
\newacronym{cots}{COTS}{Commercial Off-the-Shelf}
\newacronym{cp}{CP}{Control Plane}
\newacronym{cpu}{CPU}{Central Processing Unit}
\newacronym{cqi}{CQI}{Channel Quality Information}
\newacronym{cr}{CR}{Cognitive Radio}
\newacronym{cran}{CRAN}{Cloud \gls{ran}}
\newacronym{crs}{CRS}{Cell Reference Signal}
\newacronym{csi}{CSI}{Channel State Information}
\newacronym{csirs}{CSI-RS}{Channel State Information - Reference Signal}
\newacronym{cu}{CU}{Central Unit}
\newacronym{d2tcp}{D$^2$TCP}{Deadline-aware Data center TCP}
\newacronym{d3}{D$^3$}{Deadline-Driven Delivery}
\newacronym{dac}{DAC}{Digital to Analog Converter}
\newacronym{dag}{DAG}{Directed Acyclic Graph}
\newacronym{darpa}{DARPA}{Defense Advanced Research Projects Agency}
\newacronym{das}{DAS}{Distributed Antenna System}
\newacronym{dash}{DASH}{Dynamic Adaptive Streaming over HTTP}
\newacronym{dc}{DC}{Dual Connectivity}
\newacronym{dccp}{DCCP}{Datagram Congestion Control Protocol}
\newacronym{dce}{DCE}{Direct Code Execution}
\newacronym{dci}{DCI}{Downlink Control Information}
\newacronym{dcl}{DCL}{Dear Colleague Letter}
\newacronym{dctcp}{DCTCP}{Data Center TCP}
\newacronym{devops}{DevOps}{Development and Operations}
\newacronym{dl}{DL}{Downlink}
\newacronym{dmr}{DMR}{Deadline Miss Ratio}
\newacronym{dmrs}{DMRS}{DeModulation Reference Signal}
\newacronym{drlcc}{DRL-CC}{Deep Reinforcement Learning Congestion Control}
\newacronym{drs}{DRS}{Discovery Reference Signal}
\newacronym{dtn}{DTN}{Digital Twin Network}
\newacronym{dtmn}{DTMN}{Digital Twins for Mobile Networks}
\newacronym{dtwn}{DTWN}{Digital Twin Wireless Network}
\newacronym{du}{DU}{Distributed Unit}
\newacronym{e2e}{E2E}{end-to-end}
\newacronym{ecaas}{ECaaS}{Edge-Cloud-as-a-Service}
\newacronym{ecn}{ECN}{Explicit Congestion Notification}
\newacronym{edf}{EDF}{Earliest Deadline First}
\newacronym{em}{EM}{Electro-Magnetic}
\newacronym{embb}{eMBB}{Enhanced Mobile Broadband}
\newacronym{empower}{EMPOWER}{EMpowering transatlantic PlatfOrms for advanced WirEless Research}
\newacronym{enb}{eNB}{evolved Node Base}
\newacronym{endc}{EN-DC}{E-UTRAN-\gls{nr} \gls{dc}}
\newacronym{epc}{EPC}{Evolved Packet Core}
\newacronym{eps}{EPS}{Evolved Packet System}
\newacronym{es}{ES}{Edge Server}
\newacronym{etsi}{ETSI}{European Telecommunications Standards Institute}
\newacronym[firstplural=Estimated Times of Arrival (ETAs)]{eta}{ETA}{Estimated Time of Arrival}
\newacronym{eutran}{E-UTRAN}{Evolved Universal Terrestrial Access Network}
\newacronym{faas}{FaaS}{Function-as-a-Service}
\newacronym{fapi}{FAPI}{Functional Application Platform Interface}
\newacronym{fcc}{FCC}{Federal Communications Commission}
\newacronym{fdd}{FDD}{Frequency Division Duplexing}
\newacronym{fdm}{FDM}{Frequency Division Multiplexing}
\newacronym{fdma}{FDMA}{Frequency Division Multiple Access}
\newacronym{fed4fire}{FED4FIRE+}{Federation 4 Future Internet Research and Experimentation Plus}
\newacronym{fir}{FIR}{Finite Impulse Response}
\newacronym{fl}{FL}{Federated Learning}
\newacronym{fpga}{FPGA}{Field Programmable Gate Array}
\newacronym{fr2}{FR2}{Frequency Range 2}
\newacronym{fs}{FS}{Fast Switching}
\newacronym{fscc}{FSCC}{Flow Sharing Congestion Control}
\newacronym{ftp}{FTP}{File Transfer Protocol}
\newacronym{fw}{FW}{Flow Window}
\newacronym{ga128}{Ga}{Golay Sequence type A}
\newacronym{ge}{GE}{Gaussian Elimination}
\newacronym{glfsr}{GLFSR}{Galois Linear Feedback Shift Register}
\newacronym{gnb}{gNB}{Next Generation Node Base}
\newacronym{gold}{Gold}{Gold}
\newacronym{gop}{GOP}{Group of Pictures}
\newacronym{gpr}{GPR}{Gaussian Process Regressor}
\newacronym{gpu}{GPU}{Graphics Processing Unit}
\newacronym{gtp}{GTP}{GPRS Tunneling Protocol}
\newacronym{gtpc}{GTP-C}{GPRS Tunnelling Protocol Control Plane}
\newacronym{gtpu}{GTP-U}{GPRS Tunnelling Protocol User Plane}
\newacronym{gtpv2c}{GTPv2-C}{\gls{gtp} v2 - Control}
\newacronym{gw}{GW}{Gateway}
\newacronym{harq}{HARQ}{Hybrid Automatic Repeat reQuest}
\newacronym{hetnet}{HetNet}{Heterogeneous Network}
\newacronym{hh}{HH}{Hard Handover}
\newacronym{hol}{HOL}{Head-of-Line}
\newacronym{hqf}{HQF}{Highest-quality-first}
\newacronym{hss}{HSS}{Home Subscription Server}
\newacronym{http}{HTTP}{HyperText Transfer Protocol}
\newacronym{ia}{IA}{Initial Access}
\newacronym{iab}{IAB}{Integrated Access and Backhaul}
\newacronym{ic}{IC}{Incident Command}
\newacronym{ietf}{IETF}{Internet Engineering Task Force}
\newacronym{ifw}{IFW}{Interference Free Window}
\newacronym{imsi}{IMSI}{International Mobile Subscriber Identity}
\newacronym{imt}{IMT}{International Mobile Telecommunication}
\newacronym{iot}{IoT}{Internet of Things}
\newacronym{ip}{IP}{Internet Protocol}
\newacronym{iq}{IQ}{In-phase and Quadrature}
\newacronym{isi}{ISI}{Inter-Symbol Interference}
\newacronym{itu}{ITU}{International Telecommunication Union}
\newacronym{kpi}{KPI}{Key Performance Indicator}
\newacronym{kvm}{KVM}{Kernel-based Virtual Machine}
\newacronym{lfsr}{LFSR}{Linear Feedback Shift Register}
\newacronym{los}{LOS}{Line-of-Sight}
\newacronym{ls}{LS}{Loosely Synchronised}
\newacronym{lsm}{LSM}{Link-to-System Mapping}
\newacronym{lstm}{LSTM}{Long Short Term Memory}
\newacronym{lte}{LTE}{Long Term Evolution}
\newacronym{lxc}{LXC}{Linux Container}
\newacronym{m2m}{M2M}{Machine to Machine}
\newacronym{mac}{MAC}{Medium Access Control}
\newacronym{mai}{MAI}{Multiple Access Interference}
\newacronym{manet}{MANET}{Mobile Ad Hoc Network}
\newacronym{mano}{MANO}{Management and Orchestration}
\newacronym{mc}{MC}{Multi-Connectivity}
\newacronym{mcc}{MCC}{Mobile Cloud Computing}
\newacronym{mchem}{MCHEM}{Massive Channel Emulator}
\newacronym{mcs}{MCS}{Modulation and Coding Scheme}
\newacronym{mec}{MEC}{Multi-access Edge Computing}
\newacronym{mec2}{MEC}{Mobile Edge Cloud}
\newacronym{mec3}{MEC}{Mobile Edge Computing}
\newacronym{mfc}{MFC}{Mobile Fog Computing}
\newacronym{mi}{MI}{Mutual Information}
\newacronym{mib}{MIB}{Master Information Block}
\newacronym{miesm}{MIESM}{Mutual Information Based Effective SINR}
\newacronym{mimo}{MIMO}{Multiple Input, Multiple Output}
\newacronym{mgen}{MGEN}{Multi-Generator}
\newacronym{ml}{ML}{Machine Learning}
\newacronym{mlr}{MLR}{Maximum-local-rate}
\newacronym[plural=\gls{mme}s,firstplural=Mobility Management Entities (MMEs)]{mme}{MME}{Mobility Management Entity}
\newacronym{mmtc}{mMTC}{Massive Machine-Type Communications}
\newacronym{mmwave}{mmWave}{millimeter wave}
\newacronym{mpdccp}{MP-DCCP}{Multipath Datagram Congestion Control Protocol}
\newacronym{mptcp}{MPTCP}{Multipath TCP}
\newacronym{mr}{MR}{Maximum Rate}
\newacronym{mrdc}{MR-DC}{Multi \gls{rat} \gls{dc}}
\newacronym{mse}{MSE}{Mean Square Error}
\newacronym{mss}{MSS}{Maximum Segment Size}
\newacronym{mt}{MT}{Mobile Termination}
\newacronym{mtd}{MTD}{Machine-Type Device}
\newacronym{mtu}{MTU}{Maximum Transmission Unit}
\newacronym{mumimo}{MU-MIMO}{Multi-user \gls{mimo}}
\newacronym{mvno}{MVNO}{Mobile Virtual Network Operator}
\newacronym{nalu}{NALU}{Network Abstraction Layer Unit}
\newacronym{nas}{NAS}{Network Attached Storage}
\newacronym{nbiot}{NB-IoT}{Narrow Band IoT}
\newacronym{nfv}{NFV}{Network Function Virtualization}
\newacronym{nfvi}{NFVI}{Network Function Virtualization Infrastructure}
\newacronym{nic}{NIC}{Network Interface Card}
\newacronym{nlos}{NLOS}{Non-Line-of-Sight}
\newacronym{now}{NOW}{Non Overlapping Window}
\newacronym{nrdz}{NRDZ}{National Radio Dynamic Zone}
\newacronym{nsf}{NSF}{National Science Foundation}
\newacronym{nsm}{NSM}{Network Service Mesh}
\newacronym{nr}{NR}{New Radio}
\newacronym{nrf}{NRF}{Network Repository Function}
\newacronym{nsa}{NSA}{Non Stand Alone}
\newacronym{nse}{NSE}{Network Slicing Engine}
\newacronym{nssf}{NSSF}{Network Slice Selection Function}
\newacronym{ntp}{NTP}{Network Time Protocol}
\newacronym{o2i}{O2I}{Outdoor to Indoor}
\newacronym{oai}{OAI}{OpenAirInterface}
\newacronym{oaicn}{OAI-CN}{\gls{oai} \acrlong{cn}}
\newacronym{oairan}{OAI-RAN}{\acrlong{oai} \acrlong{ran}}
\newacronym{oam}{OAM}{Operations, Administration and Maintenance}
\newacronym[plural=\gls{obu}s,firstplural=Onboard Units (OBUs)]{obu}{OBU}{Onboard Unit}
\newacronym{ofdm}{OFDM}{Orthogonal Frequency Division Multiplexing}
\newacronym{olia}{OLIA}{Opportunistic Linked Increase Algorithm}
\newacronym{omec}{OMEC}{Open Mobile Evolved Core}
\newacronym{onap}{ONAP}{Open Network Automation Platform}
\newacronym{onf}{ONF}{Open Networking Foundation}
\newacronym{onos}{ONOS}{Open Networking Operating System}
\newacronym{oom}{OOM}{\gls{onap} Operations Manager}
\newacronym{opnfv}{OPNFV}{Open Platform for \gls{nfv}}
\newacronym{orbit}{ORBIT}{Open-Access Research Testbed for Next-Generation Wireless Networks}
\newacronym{os}{OS}{Operating System}
\newacronym{osm}{OSM}{Open Street Map}
\newacronym{oss}{OSS}{Operations Support System}
\newacronym{pa}{PA}{Position-aware}
\newacronym{pase}{PASE}{Prioritization, Arbitration, and Self-adjusting Endpoints}
\newacronym{pawr}{PAWR}{Platforms for Advanced Wireless Research}
\newacronym{pbch}{PBCH}{Physical Broadcast Channel}
\newacronym{pcef}{PCEF}{Policy and Charging Enforcement Function}
\newacronym{pcfich}{PCFICH}{Physical Control Format Indicator Channel}
\newacronym{pcrf}{PCRF}{Policy and Charging Rules Function}
\newacronym{pdcch}{PDCCH}{Physical Downlink Control Channel}
\newacronym{pdcp}{PDCP}{Packet Data Convergence Protocol}
\newacronym{pdsch}{PDSCH}{Physical Downlink Shared Channel}
\newacronym{pdu}{PDU}{Packet Data Unit}
\newacronym{pdp}{PDP}{Power Delay Profile}
\newacronym{pf}{PF}{Proportional Fair}
\newacronym{pgw}{PGW}{Packet Gateway}
\newacronym{phich}{PHICH}{Physical Hybrid ARQ Indicator Channel}
\newacronym{phy}{PHY}{Physical}
\newacronym{pl}{PL}{Path Loss}
\newacronym{pmch}{PMCH}{Physical Multicast Channel}
\newacronym{pmi}{PMI}{Precoding Matrix Indicators}
\newacronym{powder}{POWDER}{Platform for Open Wireless Data-driven Experimental Research}
\newacronym{ppo}{PPO}{Proximal Policy Optimization}
\newacronym{ppp}{PPP}{Poisson Point Process}
\newacronym{prach}{PRACH}{Physical Random Access Channel}
\newacronym{prb}{PRB}{Physical Resource Block}
\newacronym{psnr}{PSNR}{Peak Signal to Noise Ratio}
\newacronym{pss}{PSS}{Primary Synchronization Signal}
\newacronym{pucch}{PUCCH}{Physical Uplink Control Channel}
\newacronym{pusch}{PUSCH}{Physical Uplink Shared Channel}
\newacronym{qam}{QAM}{Quadrature Amplitude Modulation}
\newacronym{qci}{QCI}{\gls{qos} Class Identifier}
\newacronym{qoe}{QoE}{Quality of Experience}
\newacronym{qos}{QoS}{Quality of Service}
\newacronym{qtgui}{QT-GUI}{QT Graphical User Interface}
\newacronym{quic}{QUIC}{Quick UDP Internet Connections}
\newacronym{rach}{RACH}{Random Access Channel}
\newacronym{ran}{RAN}{Radio Access Network}
\newacronym[firstplural=Radio Access Technologies (RATs)]{rat}{RAT}{Radio Access Technology}
\newacronym{rcn}{RCN}{Research Coordination Network}
\newacronym{rec}{REC}{Radio Edge Cloud}
\newacronym{red}{RED}{Random Early Detection}
\newacronym{renew}{RENEW}{Reconfigurable Eco-system for Next-generation End-to-end Wireless}
\newacronym{rf}{RF}{Radio Frequency}
\newacronym{rfc}{RFC}{Request for Comments}
\newacronym{rfr}{RFR}{Random Forest Regressor}
\newacronym{ric}{RIC}{\gls{ran} Intelligent Controller}
\newacronym{rlc}{RLC}{Radio Link Control}
\newacronym{rlf}{RLF}{Radio Link Failure}
\newacronym{rlnc}{RLNC}{Random Linear Network Coding}
\newacronym{rmse}{RMSE}{Root Mean Squared Error}
\newacronym{rnis}{RNIS}{Radio Network Information Service}
\newacronym{rr}{RR}{Round Robin}
\newacronym{rrc}{RRC}{Radio Resource Control}
\newacronym{rrm}{RRM}{Radio Resource Management}
\newacronym{rru}{RRU}{Remote Radio Unit}
\newacronym{rs}{RS}{Remote Server}
\newacronym{rsrp}{RSRP}{Reference Signal Received Power}
\newacronym{rsrq}{RSRQ}{Reference Signal Received Quality}
\newacronym{rss}{RSS}{Received Signal Strength}
\newacronym{rssi}{RSSI}{Received Signal Strength Indicator}
\newacronym{rsu}{RSU}{Road-Side Unit}
\newacronym{rtt}{RTT}{Round Trip Time}
\newacronym{ru}{RU}{Radio Unit}
\newacronym{rw}{RW}{Receive Window}
\newacronym{rx}{RX}{Receiver}
\newacronym{s1ap}{S1AP}{S1 Application Protocol}
\newacronym{sa}{SA}{standalone}
\newacronym{sack}{SACK}{Selective Acknowledgment}
\newacronym{sap}{SAP}{Service Access Point}
\newacronym{sc2}{SC2}{Spectrum Collaboration Challenge}
\newacronym{scef}{SCEF}{Service Capability Exposure Function}
\newacronym{sch}{SCH}{Secondary Cell Handover}
\newacronym{scoot}{SCOOT}{Split Cycle Offset Optimization Technique}
\newacronym{sctp}{SCTP}{Stream Control Transmission Protocol}
\newacronym{sdap}{SDAP}{Service Data Adaptation Protocol}
\newacronym{sd}{SD}{Standard Deviation}
\newacronym{sdk}{SDK}{Software Development Kit}
\newacronym{sdm}{SDM}{Space Division Multiplexing}
\newacronym{sdma}{SDMA}{Spatial Division Multiple Access}
\newacronym{sdn}{SDN}{Software-defined Networking}
\newacronym{sdr}{SDR}{Software-defined Radio}
\newacronym{seba}{SEBA}{SDN-Enabled Broadband Access}
\newacronym{sgsn}{SGSN}{Serving GPRS Support Node}
\newacronym{sgw}{SGW}{Service Gateway}
\newacronym{si}{SI}{Study Item}
\newacronym{sib}{SIB}{Secondary Information Block}
\newacronym{sinr}{SINR}{Signal to Interference plus Noise Ratio}
\newacronym{sip}{SIP}{Session Initiation Protocol}
\newacronym{siso}{SISO}{Single Input, Single Output}
\newacronym{sla}{SLA}{Service Level Agreement}
\newacronym{sm}{SM}{Saturation Mode}
\newacronym{smf}{SMF}{Session Management Function}
\newacronym{smo}{SMO}{Service Management and Orchestration}
\newacronym{sms}{SMS}{Short Message Service}
\newacronym{smsgmsc}{SMS-GMSC}{\gls{sms}-Gateway}
\newacronym{snr}{SNR}{Signal-to-Noise-Ratio}
\newacronym{son}{SON}{Self-Organizing Network}
\newacronym{sptcp}{SPTCP}{Single Path TCP}
\newacronym{srb}{SRB}{Service Radio Bearer}
\newacronym{srn}{SRN}{Standard Radio Node}
\newacronym{srs}{SRS}{Sounding Reference Signal}
\newacronym{ss}{SS}{Synchronization Signal}
\newacronym{sss}{SSS}{Secondary Synchronization Signal}
\newacronym{st}{ST}{Spanning Tree}
\newacronym{svc}{SVC}{Scalable Video Coding}
\newacronym{tb}{TB}{Transport Block}
\newacronym{tcp}{TCP}{Transmission Control Protocol}
\newacronym{tdd}{TDD}{Time Division Duplexing}
\newacronym{tdm}{TDM}{Time Division Multiplexing}
\newacronym{tdma}{TDMA}{Time Division Multiple Access}
\newacronym{tfl}{TfL}{Transport for London}
\newacronym{tfrc}{TFRC}{TCP-Friendly Rate Control}
\newacronym{tft}{TFT}{Traffic Flow Template}
\newacronym{tgen}{TGEN}{Traffic Generator}
\newacronym{tip}{TIP}{Telecom Infra Project}
\newacronym{tm}{TM}{Transparent Mode}
\newacronym{to}{TO}{Telco Operator}
\newacronym{toa}{ToA}{Time of Arrival}
\newacronym{tr}{TR}{Technical Report}
\newacronym{trp}{TRP}{Transmitter Receiver Pair}
\newacronym{ts}{TS}{Technical Specification}
\newacronym{tti}{TTI}{Transmission Time Interval}
\newacronym{ttt}{TTT}{Time-to-Trigger}
\newacronym{tx}{TX}{Transmitter}
\newacronym{uas}{UAS}{Unmanned Aerial System}
\newacronym{uav}{UAV}{Unmanned Aerial Vehicle}
\newacronym{udm}{UDM}{Unified Data Management}
\newacronym{udp}{UDP}{User Datagram Protocol}
\newacronym{udr}{UDR}{Unified Data Repository}
\newacronym{ue}{UE}{User Equipment}
\newacronym{uhd}{UHD}{\gls{usrp} Hardware Driver}
\newacronym{ul}{UL}{Uplink}
\newacronym{um}{UM}{Unacknowledged Mode}
\newacronym{uml}{UML}{Unified Modeling Language}
\newacronym{upa}{UPA}{Uniform Planar Array}
\newacronym{upf}{UPF}{User Plane Function}
\newacronym{urllc}{URLLC}{Ultra Reliable and Low Latency Communication}
\newacronym{usa}{U.S.}{United States}
\newacronym{usim}{USIM}{Universal Subscriber Identity Module}
\newacronym{usrp}{USRP}{Universal Software Radio Peripheral}
\newacronym{utc}{UTC}{Urban Traffic Control}
\newacronym{vim}{VIM}{Virtualization Infrastructure Manager}
\newacronym{vm}{VM}{Virtual Machine}
\newacronym{vnf}{VNF}{Virtual Network Function}
\newacronym{volte}{VoLTE}{Voice over \gls{lte}}
\newacronym{voltha}{VOLTHA}{Virtual OLT HArdware Abstraction}
\newacronym{vr}{VR}{Virtual Reality}
\newacronym{vran}{vRAN}{Virtualized \gls{ran}}
\newacronym{vss}{VSS}{Video Streaming Server}
\newacronym{wbf}{WBF}{Wired Bias Function}
\newacronym{wf}{WF}{Wired-first}
\newacronym{wi}{WI}{Wireless InSite}
\newacronym{wlan}{WLAN}{Wireless Local Area Network}
\newacronym{pnf}{PNF}{Physical Network Function}
\newacronym{drl}{DRL}{Deep Reinforcement Learning}
\newacronym{mtc}{MTC}{Machine-type Communications}
\newacronym{v2x}{V2X}{Vehicle-to-everything}
\newacronym{cast}{CaST}{Channel emulation scenario generator and Sounder Toolchain}
\newacronym{gui}{GUI}{Graphical User Interface}
\newacronym{ups}{UPS}{Uninterruptible Power Supply}
\newacronym{ota}{OTA}{Over-the-Air}
\newacronym{hitl}{HITL}{hardware-in-the-loop}
\newacronym{soc}{SoC}{System-on-Chip}

\newacronym{psr}{PSR}{Packet Success Rate}
\newacronym{sadr}{SADR}{Safe Adaptive Data Rate}
\newacronym{cnn}{CNN}{Convolutional Neural Network}
\newacronym{kpm}{KPM}{Key Performance Measurement}
\newacronym{mqtt}{MQTT}{Message Queuing Telemetry Transport}
\newacronym{coap}{CoAP}{Constrained Application Protocol}
\newacronym{amqp}{AMQP}{Advanced Message Queuing Protocol}
\newacronym{dt}{DT}{Digital Twin}
\newacronym{dlai}{DL}{Deep Learning}

\def\BibTeX{{\rm B\kern-.05em{\sc i\kern-.025em b}\kern-.08em
   T\kern-.1667em\lower.7ex\hbox{E}\kern-.125emX}}
\usepackage[numbers,sort&compress]{natbib}
\usepackage{booktabs} % For better-looking tables
\usepackage{tabularx} % For tables with flexible column widths

\newcommand{\twinet}{\textit{TwiNet}\xspace}

\usepackage{tikz}
\usepackage{tikzscale}
\newif\ifexttikz
\exttikzfalse
\ifexttikz
\else
\usepackage{tikzpagenodes,etoolbox}
\usetikzlibrary{calc}
\usepackage[contents={}]{background}
\AddEverypageHook{%
\ifnumequal{\thepage}{1}{%
    \tikz[remember picture,overlay]{%
        \node[draw,
        minimum width=1.03\textwidth,
        text width=1.02\textwidth,
        font=\footnotesize
        ]
        at ($(current page header area) - (0,5pt)$)
        {%
        This paper has been accepted for publication on IEEE Global Communications Conference (GLOBECOM 2024). This is the author's accepted version of the article. The final version published by IEEE is CP. Robinson, A. Lacava, P. Johari, F. Cuomo, and T. Melodia, ``\twinet: Connecting Real World Networks to their Digital Twins Through a Live Bidirectional Link,'' Proc. of  \textit{IEEE Global Communications Conference (GLOBECOM)}, Cape Town, South Africa, 2024.
        };
        \node[draw,
        minimum width=1.03\textwidth,
        text width=1.02\textwidth,
        font=\footnotesize
        ]
        at (current page footer area)
        {%
        ©2024 IEEE. Personal use of this material is permitted. Permission from IEEE must be obtained for all other uses, in any current or future media, including reprinting/republishing this material for advertising or promotional purposes, creating new collective works, for resale or redistribution to servers or lists, or reuse of any copyrighted component of this work in other works.
        };
    }%
}{}%end ifnumequal
}
\fi

\begin{document}

\title{\twinet: Connecting Real World Networks to their Digital Twins Through a Live Bidirectional Link}  %% MQDT

\author{\IEEEauthorblockN{Clifton Paul Robinson\IEEEauthorrefmark{1}, Andrea Lacava\IEEEauthorrefmark{1}\IEEEauthorrefmark{2}, Pedram Johari\IEEEauthorrefmark{1},
Francesca Cuomo\IEEEauthorrefmark{2}, Tommaso Melodia\IEEEauthorrefmark{1}}
\IEEEauthorblockN{\IEEEauthorrefmark{1}Institute for the Wireless Internet of Things, Northeastern University, Boston, MA, USA\\
\IEEEauthorrefmark{2}Sapienza, University of Rome, Rome, Italy\\
Email: $\{$robinson.c, lacava.a,  t.melodia$\}$@northeastern.edu\\
Email: francesca.cuomo@uniroma1.it}

\vspace{-25pt}

\thanks{This work was partially supported by the National Telecommunications and Information Administration (NTIA)’s Public Wireless Supply Chain Innovation Fund (PWSCIF) under Award No. 25-60-IF011.}
%  and by the U.S. National Science Foundation under grant CNS-1925601
}

% \pagestyle{plain}
% \pagenumbering{gobble}

\maketitle

\glsresetall

\begin{abstract}

The wireless spectrum's increasing complexity poses challenges and opportunities, highlighting the necessity for real-time solutions and robust data processing capabilities. \gls{dt}, virtual replicas of physical systems, integrate real-time data to mirror their real-world counterparts, enabling precise monitoring and optimization. Incorporating \glspl{dt} into wireless communication enhances predictive maintenance, resource allocation, and troubleshooting, thus bolstering network reliability.
Our paper introduces \twinet, enabling bidirectional, near-real-time links between real-world wireless spectrum scenarios and \gls{dt} replicas. Utilizing the protocol, MQTT, we can achieve data transfer times with an average latency of 14~ms, suitable for real-time communication. This is confirmed by monitoring real-world traffic and mirroring it in real-time within the \gls{dt}'s wireless environment. We evaluate \twinet's performance in two distinct use cases: (i) enhancing \gls{sadr} systems by assessing risky traffic configurations of UEs, resulting in approximately 15\% improved network performance compared to original network selections; and (ii) deploying new CNNs in response to jammed pilots, where the DL pipeline achieves up to 97\% accuracy by training on artificial data and deploying a new model in as low as 2 minutes to counter persistent adversaries. \twinet enables swift deployment and adaptation of \glspl{dt}, addressing crucial challenges in modern wireless communication systems.

% Our paper introduces \twinet, enabling bidirectional, near-real-time links between real-world wireless spectrum scenarios and \gls{dt} replicas. Utilizing the protocol, MQTT, we can achieve data transfer times with an average latency of 14~ms, suitable for real-time communication. This is confirmed by monitoring real-world traffic and mirroring it in real-time within the \gls{dt}'s wireless environment. We evaluate \twinet's performance in two use cases: (i) assessing risky traffic configurations of UEs in a \gls{sadr} system, improving network performance by approximately 15\% compared to original network selections; and (ii) deploying new CNNs in response to jammed pilots, achieving up to 97\% accuracy training on artificial data and deploying a new model in as low as 2 minutes to counter persistent adversaries. \twinet enables swift deployment and adaptation of \glspl{dt}, addressing crucial challenges in modern wireless communication systems.

\end{abstract}

\begin{IEEEkeywords}
digital twin, wireless spectrum, testbed implementation, deep learning, network management
\end{IEEEkeywords}

\glsresetall

%%%%%%%%%%%%%%%%%%%%%%%%%%%%%%%%%%%%%%%%%%%%%%%%%%%%%%%%%%%%%%%%%%%%%%%%%%%%%%%%%%%%%%%%

\section{Introduction}
\label{sec:intro}

%As the world becomes more dependent on advanced wireless technologies, the need for robust real-time solutions grows. Technologies like \gls{iot} aim to revolutionize everyday life, but they also pose challenges. \gls{iot} devices prioritize affordability and simplicity, often lacking local processing power or storage~\cite{IBMDigitalTwinsIoT}. With a focus on cost efficiency per device, one must look to other means to solve these emerging challenges in near real-time.
%
In recent years~\cite{IBMDigitalTwin}, researchers have begun to look away from the physical world and toward the virtual one. By being able to offload data for decision-making and predictive analytics, devices with hardware limitations can gain better capabilities without significant adjustments~\cite{IBMDigitalTwinsIoT}.
One of these methods is called a \gls{dt}, a virtual model designed to accurately mirror a physical object or system. For a virtual model to be considered a \gls{dt}, an automated bidirectional communication link must connect both realms, facilitating continuous data exchange to aid in decision-making~\cite{IBMDigitalTwin}.

\begin{figure}[!t]
    \centering
    \includegraphics[width=.95\linewidth]{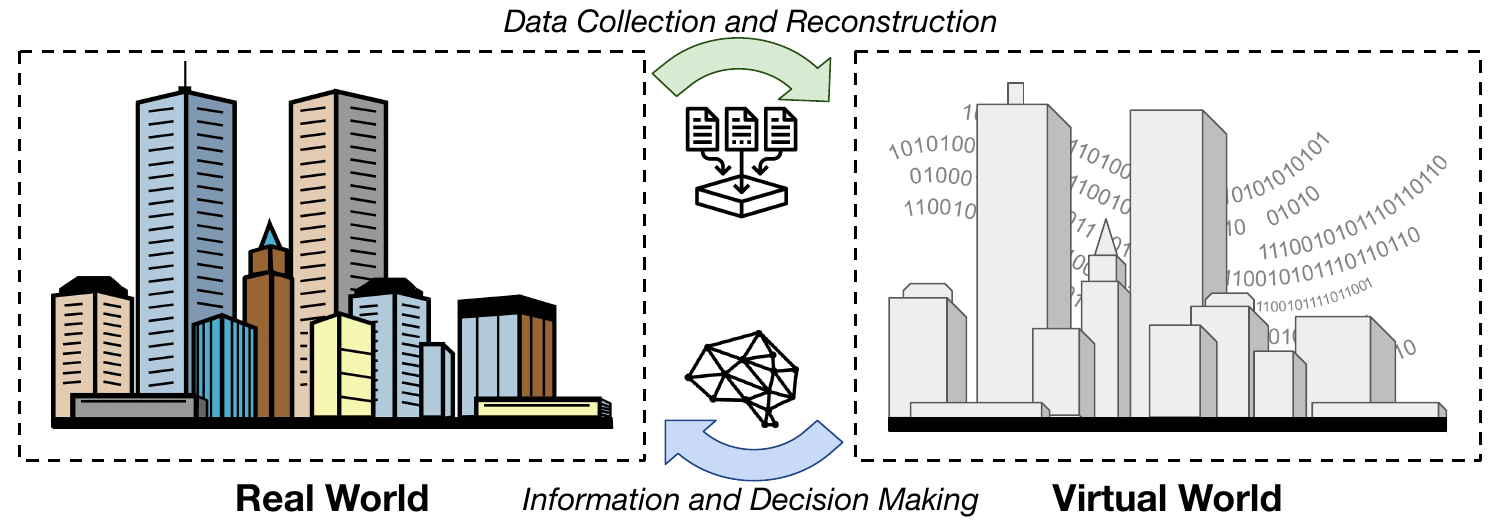}
    \caption{High-level representation of a digital twin for a smart city.}
    \label{fig:twin_intro}
    \vspace{-0.3cm}
\end{figure}

Smart cities often utilize \gls{dt} technology to improve data quality~\cite{farsi_digital_2020}. \glspl{dt} enable real-time simulations and analytics of urban systems, aiding in decision-making, resource optimization, and predictive maintenance~\cite{farsi_digital_2020}. As shown in Fig.~\ref{fig:twin_intro}, real-world data is exchanged with \glspl{dt} to support informed decision-making.
\glspl{dt} offer advantages over real-world testing, including improved efficiency via real-world mirroring for ongoing performance optimization, and near real-time communication through constant updates to closely mirror real-world counterparts for enhanced research, design, and evaluation of novel solutions in a risk-free environment, ~\cite{IBMDigitalTwin, IBMDigitalTwinsIoT}.

\begin{figure*}[!t]
	\centering
	\includegraphics[width=.95\textwidth]{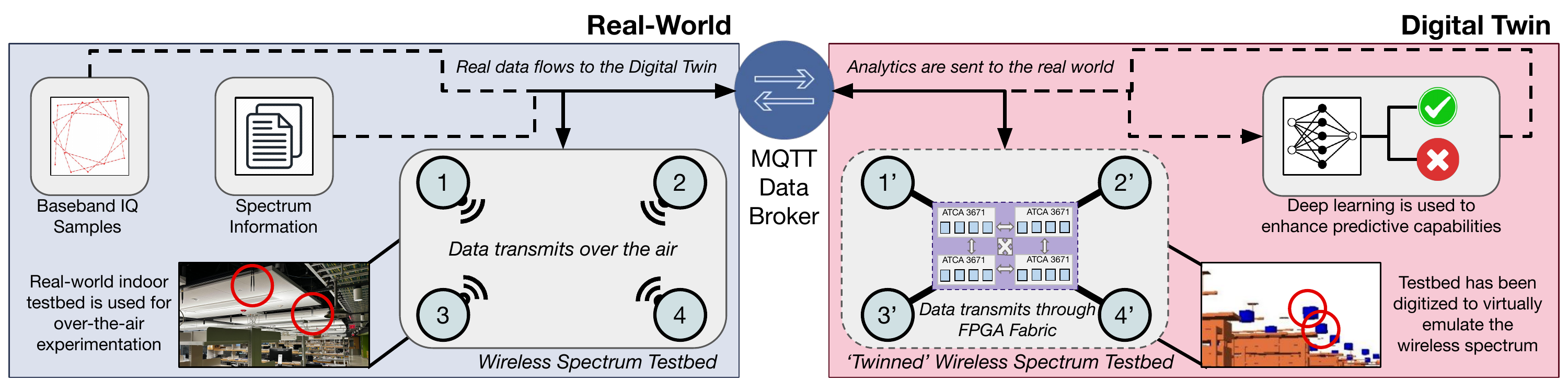}
	\caption{High-level overview of \twinet, showing the link between the real world and \gls{dt} along with the spectrum and deep learning capabilities.}
	\label{fig:twinet}
        \vspace{-0.5cm}
\end{figure*}

% From a wireless communication standpoint, ...
Integrating \glspl{dt} in wireless communication revolutionizes network efficiency, reliability, and optimization. Traditional methods lag behind the exponential growth of wireless devices and system complexity. Virtual replicas of physical network components, created through \glspl{dt}, enable real-time monitoring, predictive analysis, and scenario emulation. This approach addresses challenges, optimizes performance, and enhances network resilience. By accurately mirroring wireless spectrum behavior, \glspl{dt} identify issues, optimize resource allocation, and improve service quality. Implementing machine learning and \gls{ai}, they automate decision-making and adaptively optimize network configurations. Seamless setup and optimization of connections between real and virtual worlds are crucial for realizing their full potential.

In this paper, we introduce \twinet, a generalized approach for synchronizing novel wireless spectrum scenarios between a \gls{dt} and its real-world counterpart. A fast, bidirectional communication link is crucial for successful \gls{dt} operation, facilitating seamless data exchange. We establish this link by leveraging the \gls{iot} \gls{mqtt} protocol, emphasizing its publish/subscribe method and reliance on a single data broker. The asynchronous nature of \gls{mqtt} enhances scalability and flexibility, owing to its distinctive publishing method.
Fig.~\ref{fig:twinet} outlines \twinet's architecture and its utilization of \gls{mqtt} in various \gls{dt} scenarios. It facilitates flexible management across the \gls{dt}, enabling task allocation. Two configurations are depicted: wireless spectrum replication and \gls{dlai}-based predictive analysis, each with specific data requirements. \gls{mqtt} optimizes data transmission, targeting relevant \gls{dt} components, highlighting scalability and adaptability crucial for large-scale deployment.

% This method works extremely well for bidirectional communication in \gls{dt} systems due to its asynchronous nature, allowing real-time data exchange between the physical system and its digital counterpart without requiring both to be active simultaneously. Additionally, its decoupled architecture enables scalability and flexibility, accommodating diverse data sources and subscribers efficiently (as seen in Fig.~\ref{fig:twinet}). We prove this by testing it against well-known questions in the wireless spectrum, such as network traffic monitoring, risky states in Open RAN network performance control, and signal classification through \gls{dlai}.

% The contributions of the work should be clarified further.
The main contributions of this paper are as follows:
\begin{enumerate}
    \item We introduce \twinet, a bidirectional communication link between real-world scenarios and its \gls{dt}, achieving data transfer latencies as low as 14~ms per packet, and demonstrate real-time traffic monitoring and mirroring experiments between the real-world and its \gls{dt} counterpart.
    \item In the first use case, we implement a \gls{sadr} system that exploits the capabilities of the \gls{dt} to evaluate traffic requests from the \glspl{ue} to identify and prevent risky actions and states that can lead to outages, improving the performances of the real network.
    \item In the second use case, we implement a \gls{dlai} pipeline that allows \glspl{dt} to create new models to protect pilot carriers for base stations that do not have the resources to recreate models on the fly.
\end{enumerate}
% The main contributions of this paper are as follows:
% \begin{enumerate}
%     \item We introduce \twinet, a bidirectional communication link between real-world scenarios and its \gls{dt}, achieving data transfer latencies as low as 14~ms per packet, and demonstrate real-time traffic monitoring and mirroring experiments between the real-world and its \gls{dt} counterpart.
%     %
%     \item We implement a \gls{sadr} system that exploits the capabilities of the \gls{dt} to evaluate traffic requests from the \glspl{ue} to identify and prevent risky actions and states that can lead to outages, improving the performances of the real network.
%     %
%     \item We implement a \gls{dlai} pipeline that allows \glspl{dt} to create new models to protect pilot carriers for base stations that do not have the resources to recreate models on the fly.
% \end{enumerate}

The remainder of the paper goes as follows: Section~\ref{sec:relate} reviews prior work relevant to the paper's scope. Section~\ref{sec:dt_tb} elaborates on \twinet and our testbed setup, presenting our traffic mirroring proof-of-concept and findings. Section~\ref{sec:experiment_setup} delineates setups for risky state testing and adversarial \gls{dlai} classifier for pilot jamming. Section~\ref{sec:results} presents experimental findings. Lastly, Section~\ref{sec:conclusion} summarizes the paper's conclusions.

%%%%%%%%%%%%%%%%%%%%%%%%%%%%%%%%%%%%%%%%%%%%%%%%%%%%%%%%%%%%%%%%%%%%%%%%%%%%%%%%%%%%%%%%

\section{Related Work}
\label{sec:relate}

% ~\cite{xia_study_2022, cui_physical_2023, farsi_digital_2020}

\glspl{dt} have become a popular research area in the past several years. 
More recently, we have seen \gls{dt} research start to turn towards wireless communication as well~\cite{yang_joint_2023, villa_colosseum_2023}.
Within this topic, the three areas that stand out the most are wireless channel emulation, large-scale emulation deployments, and the messaging protocols to connect the real world to a \gls{dt}.
% For recent surveys on the topic, the reader can refer to~\cite{botin-sanabria_digital_2022, khan_digital_2022}.

% A core component of \glspl{dt} is emulation.
%
Wireless channel emulation, especially in \gls{dt}, replicates system software and hardware traits, boosting performance. Huang et al.~\cite{huang_towards_2020} focus on emulating massive MIMO channels, tackling scaling and accuracy challenges in channel matrix modeling. They analyze fidelity versus resource tradeoffs, concluding that FPGA-based emulation notably boosts implementation efficiency.
%
% Large-scale emulation implementations are explored in~\cite{matsumura_development_2022}, which focuses on large-scale wireless emulation in 5G, leveraging 5G New Radio technology and proposing emulation techniques for NextG technology.
%
For large-scale wireless emulation testbeds, Colosseum, an open-access and publicly available option, offers scalable FPGA architecture for real-time RF channel emulation. Researchers can tailor custom channel emulators to their hardware requirements, RF bandwidth, channel fidelity, and antenna count. Colosseum supports diverse deployments and channel conditions, encompassing cellular and Wi-Fi technologies~\cite{9677430}.
% , tehrani-moayyed_creating_2021

For wireless communication \glspl{dt}, current research addresses communication-assisted sensing~\cite{yang_joint_2023}, broad AI techniques~\cite{yang_joint_2023}, and NextG technologies~\cite{villa_colosseum_2023}.
\cite{yang_joint_2023} investigates low-latency communication and computation resource allocation in wireless network \glspl{dt}, demonstrating up to 51\% reduction in transmission delay compared to other schemes. 
Villa et al.~\cite{villa_colosseum_2023} implement the largest \gls{dt} in wireless network emulation. Through experimentation on both real and \gls{dt} instances, the authors validate the efficacy of the \gls{dt} approach, particularly in large-scale wireless systems such as Colosseum. This demonstration underscores the \gls{dt}'s potential in ensuring the precision and reliability of emulation platforms for NextG wireless networks.

Messaging protocols connect the real and \gls{dt} worlds.
In~\cite{dt_mqtt_comp}, the authors compare MQTT, CoAP, and AMQP for \gls{dt} implementations. They find that for applications prioritizing reliable delivery, such as handling large data chunks, MQTT without specific \gls{qos} settings strikes an optimal balance between latency and data transmission volume. Conversely, when reliability is less crucial, CoAP becomes the preferred choice.
Whereas in \cite{col_oran_dt}, the authors look into how packet size and network constraints impact the latency of transmissions between the real world and its \gls{dt}, showing that while MQTT is a reliable protocol, latency will be added from the use of a broker and general network congestion can add up to 10~ms in added latency.
\cite{dt_mqtt_pipeline} delves into the use of \glspl{dt} in Industry 4.0, emphasizing the Six-Layer Architecture for Digital Twins with Aggregation (SLADTA). It assesses \gls{mqtt} as a communication protocol for SLADTA-based \glspl{dt} via a case study on a simulated heliostat field, showcasing \gls{mqtt}'s efficacy for complex system \glspl{dt} within SLADTA.

This paper differs from previous studies by focusing on the implementation of \gls{dt} rather than specific applications. While Villa et al.~\cite{villa_colosseum_2023} compare test outcomes of real-world testbeds and their \gls{dt}, we highlight the real-time synchronization between real-world systems and their \gls{dt}. 
While in \cite{yang_joint_2023}, the authors focus on a single device being twinned in simulation, where we twin the spectrum itself through emulation.
Ensuring alignment between accurate data and experimental results is crucial for \gls{dt}, as is maintaining a consistent data flow between the two. we aim to develop a comprehensive framework for wireless network \gls{dt}, contrasting with experimental-focused methodologies.

\begin{table}[!b]
\setlength\belowcaptionskip{5pt}
    \vspace{-0.38cm}
    \centering
    \footnotesize
    \setlength{\tabcolsep}{2pt}
    \caption{Packet arrival times between the \gls{dt} and the real-world counterpart in milliseconds~(ms). \textit{(Averaged latency of 100 samples)}}
    \label{tab:mqtt_dt}
    \begin{tabularx}{\columnwidth}{
        >{\raggedright\arraybackslash\hsize=0.8\hsize}X 
        >{\centering\arraybackslash\hsize=1\hsize}X
        >{\centering\arraybackslash\hsize=1\hsize}X }
        \toprule
        Packet Size & Real-to-Twin Latency & Twin-to-Real Latency \\
        \midrule
        1 Byte          & 15.32~ms     & 15.12~ms \\
        100 Bytes       & 15.55~ms     & 15.69~ms \\
        1 Kilobyte      & 15.74~ms     & 15.71~ms \\
        10 Kilobytes    & 22.26~ms     & 21.98~ms \\
        100 Kilobytes   & 31.10~ms     & 31.10~ms \\
        1 Megabyte      & 47.97~ms     & 48.00~ms \\
        \bottomrule
    \end{tabularx}
    % \vspace{-10pt}
\end{table}

\section{Our Testbed Setup}
\label{sec:dt_tb}

%% Overview (Colosseum and \gls{mqtt})
\gls{dt} implementations have three primary requirements: 
\textbf{(1)} There must be a virtual model of a physical object or location that accurately reflects the real world.
\textbf{(2)} A near real-time automated bidirectional communication link must be in place to ensure continuous communication.
And \textbf{(3)} the capability to run large-scale experiments, allowing for multiple processes to run simultaneously.
To meet these requirements, we utilize Colosseum~\cite{9677430}, the world's largest network emulator, and the Arena wireless testbed~\cite{bertizzolo2020arena}. 
We implement a link using the \gls{mqtt} protocol to bridge the two together. By employing these components together, we can set up and demonstrate a full \gls{dt} implementation of the Arena testbed.

% Hardware Specifications

We deploy our \gls{dt} on the Colosseum wireless network emulator, providing a publicly accessible platform for large-scale experiments. With a capacity for 128 programmable nodes and radios, Colosseum enables evaluations within a cellular network context, mirroring real-world scenarios.
%We deploy our \gls{dt} on the Colosseum wireless network emulator, offering a publicly accessible platform for conducting large-scale experiments. With support for up to 128 programmable nodes and radio devices, Colosseum enables evaluations at scale within a softwarized cellular network context. This testbed facilitates assessments across various wireless scenarios, mirroring real-world urban cellular deployments, channels, and traffic demands.
%
Leveraging the Arena testbed scenario within Colosseum, we conduct over-the-air and emulated \gls{dt} experiments (shown in Fig.~\ref{fig:twinet}). Arena, an indoor office testbed, is versatile due to experimentation in synchronized multi-cell 5G networks, AI-powered RF, secure wireless communications, cognitive radio spectrum sensing, and numerous other areas~\cite{bertizzolo2020arena}.
%
% Again, Fig.~\ref{fig:twinet} provides a high-level overview of \twinet, illustrating the consistency in setup between the real-world and twinned testbeds, with the former communicating over-the-air and the latter employing emulation via the FPGA.

% 'Real-time' \gls{mqtt} Link
We opted for the \gls{mqtt} protocol based on findings from~\cite{dt_mqtt_comp} and \cite{cliff_lanman_demo}, prioritizing reliable packet delivery in large-scale environments. To address latency concerns, we fine-tune data transmission to send only necessary data over the \gls{dt} link, tailored to each experiment's needs.
%We chose the \gls{mqtt} protocol for our live network link due to the result from~\cite{dt_mqtt_comp}. Since we require reliable packet delivery in large-scale environments, \gls{mqtt} offered the best implementation. To mitigate the potential latency issues, we optimize data transmission methods, ensuring only required data is sent over the \gls{dt} link, which is experiment-specific.
%
%Renowned for its lightweight efficiency, reliable delivery, scalability, and bi-directional communication, \gls{mqtt} is widely employed in \gls{iot} applications. Despite potential latency trade-offs, our implementation leverages \gls{mqtt}'s advantages crucial for effective \gls{dt}. To mitigate added latency, we optimize data transmission methods.
%
To depict system communication, Table~\ref{tab:mqtt_dt} displays how packet size affects latency. Smaller packets yield lower latency, suitable for sensor readings or simple commands. 
Packet sizes up to 100\,bytes demonstrate comparable latency values, ideal for brief message exchanges.
Larger packets accommodate substantial data transfers, such as spectrum information, while maintaining acceptable latency. 
Given our \gls{dt} experiments require less than 10\,kB of data to update our experiments, our average latency time stays under 20\,ms.
\newline
%Continuous Traffic Mirroring/Proof-of-Concept
% \begin{figure}[!t]
%     \centering
%     \includegraphics[width=\linewidth]{figs/mgen_experiment.pdf}
%     \caption{Example of the MGEN traffic monitoring experiment where the real-world network updates the \gls{dt} in real-time.}
%     \label{fig:mqtt-col-arena}
%     \vspace{-0.3cm}
% \end{figure}
\textbf{Proof-of-Concept:} Real-Time Traffic Monitoring.

To validate our \twinet \gls{mqtt} connection, we conduct a proof of concept experiment using the Multi-Generator (MGEN) Network Test Tool~\cite{mgen}. This experiment involves two nodes, a transmitter, and a receiver, continuously transmitting traffic over a set duration. 
The \gls{dt} provides real-time updates on network status, reflecting changes as they occur. 
We assess the \gls{dt}'s synchronization with the real traffic patterns by measuring the time it takes for the traffic to update within the \gls{dt}, evaluating the accuracy and delay. 
The utilization of the MGEN tool is key due to its established reliability and precision, ensuring the validity and robustness of experimental results.

\begin{figure}[!t]
    %\vspace{-0.25cm}
    \centering
    \includegraphics[width=\linewidth]{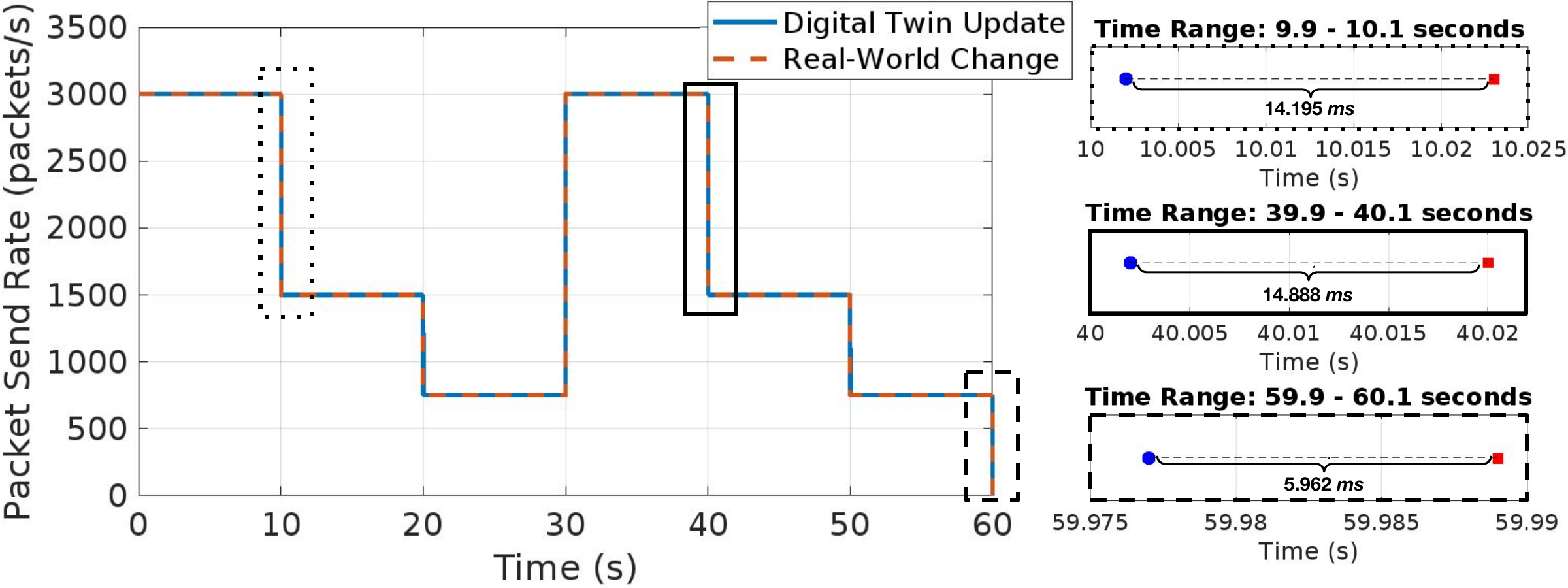}
    \caption{Results of the real-time traffic monitoring over a 60-second experiment (LEFT); Zoomed-in graphs to show the time delay on the \gls{dt} (RIGHT)}
    \label{fig:mgen-exp}
    \vspace{-0.3cm}
\end{figure}

Our experiment assesses the delay within the \gls{dt} via traffic mirroring, spanning 60\,seconds, and incorporating six alterations to the transmitter's data rate, corresponding to the number of packets sent per second in MGEN.
It is important to note that since \glspl{dt} are autonomous, the only human interference is initiating the test; everything else runs autonomously.
The averaged results of the traffic monitoring experiment are presented in Fig.~\ref{fig:mgen-exp}. Zooming into the graph is necessary due to potential overlap in results to discern the temporal discrepancy between real-world changes and the \gls{dt}'s update intervals. 
Despite some small fluctuations, the average delay between packet size changes is 13.66\,milliseconds, showing that the \gls{dt} can monitor traffic in near real-time with the \gls{mqtt} connection.

%%%%%%%%%%%%%%%%%%%%%%%%%%%%%%%%%%%%%%%%%%%%%%%%%%%%%%%%%%%%%%%%%%%%%%%%%%%%%%%%%%%%%%%%

\section{Experimental Setups}
\label{sec:experiment_setup}

To assess \twinet, we perform two distinct experiments: \textbf{(1)} utilizing \gls{sadr} for traffic management; and \textbf{(2)} addressing pilot carrier jamming mitigation via \gls{dlai}-enabled spectrum sensing.

\glslocalreset{sadr}
\subsection{\gls{sadr} for Traffic Management}

One reason for the interest in developing \glspl{dt} is their utility in environments deemed risky, in cellular networks, particularly within the intelligent control loops outlined by the Open RAN standard~\cite{bonati2021intelligence}.
Risky actions and states impacting network performance control can be observed in these settings. Risky actions involve configurations chosen by network controllers, including control heuristics or \gls{ai} algorithms, which may lead to network malfunctions due to control traffic variables misuse. 
Risky states, however, stem from unpredictable factors like network events or resource demands, causing outages for end-users. Risky actions can be anticipated, while risky states emerge unexpectedly within network traffic or radio channels.

%both risky actions and risky states can be discerned in the context of network performance control. Risky actions entail configurations selected by a generic network controller, such as a control heuristic or an \gls{ai} control algorithm, which could potentially lead to network malfunction due to the misuse of control traffic variables. Conversely, risky states are characterized as specific conditions within network traffic or the radio channel where resource demand, network event circumstances, or inappropriate configurations culminate in outage situations for end-users. Although both result in outages, risky actions can be partially identified beforehand, whereas risky states arise from conditions that cannot be fully identified or predicted in advance.

% \glslocalreset{sadr}
In this work, we explore risky action identification through \twinet, bridging the Arena real-world deployment~\cite{bertizzolo2020arena} and its \gls{dt} in Colosseum~\cite{9677430}. We implement a reference scenario as a \gls{sadr} for private \gls{lte} deployment, based on the Scope framework~\cite{bonati2021scope}, at both ends of our \gls{dt}. The term "Safe" denotes the primary objective of the ADR, aiming to minimize outage impact resulting from risky actions while optimizing data rate allocation for \glspl{ue}.

\begin{figure}[!t]
    %\vspace{0.02cm}
    \centering
    \includegraphics[width=.92\linewidth]{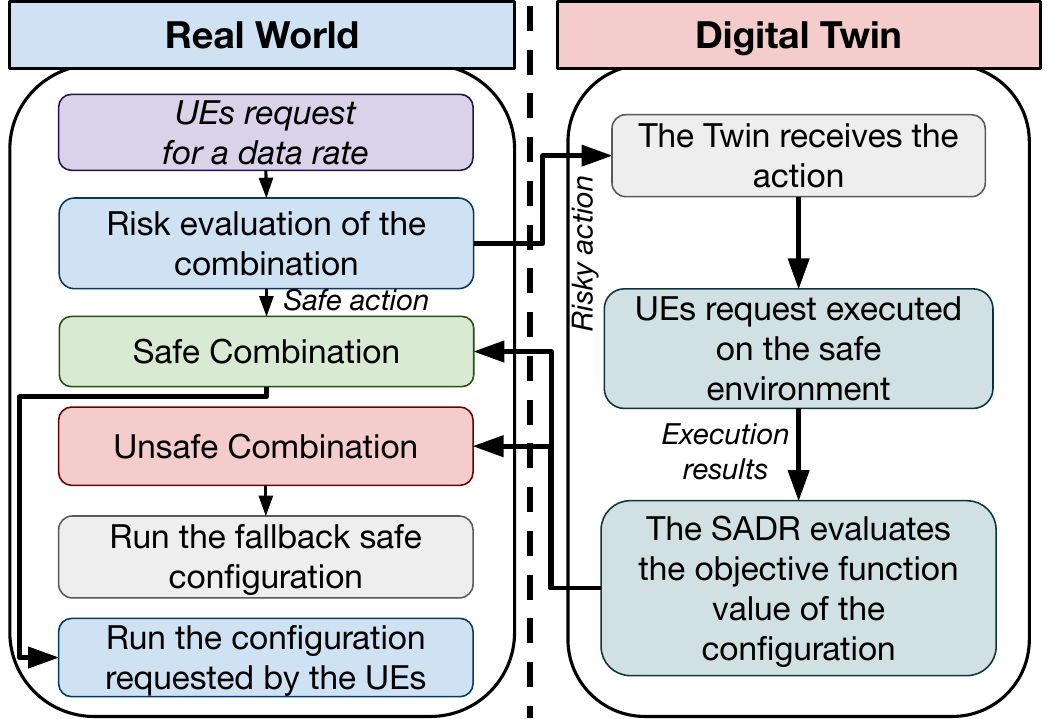}
    \caption{The \gls{sadr} system's data exchange with the real world and the \gls{dt} is visually represented via continuous data flow facilitated by \twinet.}
    \label{fig:sadr_system}
    \vspace{-0.3cm}
\end{figure}

In this deployment, $n = 3$ \glspl{ue} can simultaneously request downlink traffic from the \gls{lte} \gls{bs}. 
The \gls{sadr} aims to optimize traffic coverage while minimizing packet loss in scenarios such as critical and tactical environments where direct intervention in radio channel conditions or resource management is impractical. However, messages need to reach users accurately and promptly with minimal need for retransmission.

Such a scenario has two main goals: interconnecting a real \gls{sdr}-based \gls{ran} deployment with a virtual cellular network counterpart through \twinet, and studying the Optimization Network Problem~(\ref{eqn:sadr_problem_formulation}).
In this problem, our objective is to maximize the cumulative sum of the difference between the \gls{psr} $PSR_{i}(t)$ and the compliance between the data rate expected by the \glspl{ue} in the real world. 
$r^{EXP}_i(t)$ and the data rate granted by the \gls{sadr} $r^{ACT}_i(t)$ to minimize the risk of outages.
The \gls{sadr} can control the system with a control variable $\mathbf{a} = \{a_i(t)\}_{i=1,\dots,N,~t=0,\dots, +\infty}$ that represents one choice from the set of actions $A$ with $|A| = 10$ possible exclusive traffic combinations for the \glspl{ue} that can allocated at a given time $t$, ranging from no traffic up to 4.5\,Mbps. 
Such actions are then mapped to a specific $r^{EXP}_i$ by a function $f : \mathbb{R} \Rightarrow A$. The combination of such requests from the \glspl{ue} can represent risky actions within the system:

\vspace{-0.4cm}
\begin{align}
\max_{\mathbf{a}} & \hspace{0.2cm} \sum_{t=t_0}^{\infty}\sum_{i=1}^{N} PSR_{i}(t) - \frac{r^{EXP}_i(t) - r^{ACT}_i(t)}{r^{EXP}_i(t)} \label{eqn:sadr_problem_formulation} \tag{ONP}\\
\text{subject to} & \hspace{0.2cm} r^{EXP}_i(t) = f (a_i(t)) \hspace{0.15cm} \forall i=1,\dots, N \label{eqn:sadr_form1}\\
    & \hspace{0.2cm} a_i(t) \in A, \hspace{1.55cm} \forall t=0,\dots, +\infty
\label{eqn:sadr_form2}
\end{align}

In these conditions, the \gls{sadr} assesses risks by relaying proposed actions to its \gls{dt} via \twinet, predicting outages, and adjusting data rates for each \gls{ue} to maintain network stability.
% Under these conditions, the \gls{sadr} assesses risks by transmitting the proposed action to its \gls{dt} counterpart via \twinet. This assessment predicts potential outages, prompting the \gls{sadr} to modify the requested data rate from each \gls{ue} to maintain network conditions.

For the \gls{sadr} problem, we define the objective function as the cumulative difference between the requested data rate $r^{EXP}_i(t)$ and the actual data rate $r^{ACT}_i(t)$ for the i-th \gls{ue}, controlled by the \gls{sadr} in real-world configurations. Each \gls{ue} has 10\,possible traffic alternatives ranging from no traffic to 4.5\,Mbps, denoted by set $A$.
The \gls{sadr} assesses the risk associated with these requests through \twinet, anticipating potential outages and adjusting data rates accordingly. We identify risky bandwidth combinations based on the implicit causal relation between requested bandwidth and outage likelihood under constant resource availability and similar channel conditions.
Algorithm~\ref{alg:sadr} presents the \gls{sadr} control decision routine, employing \twinet to analyze requests and determine whether to grant them or decrease bandwidth allocation. This routine comprises two main functions: one triggered by new traffic requests from \glspl{ue}, and the other by the \gls{dt} finishing evaluation of risky requests.
Initially, the \gls{sadr} assesses the risk of requests and forwards them to the \gls{dt} for execution in a controlled environment, segmenting actions based on their impact on radio resources and requested data rates.

The \gls{sadr} assesses the \gls{dt} results to determine if the configuration meets requirements or if \glspl{ue} data rates need adjustment. The minimum reward safety threshold for \glspl{ue} is based on scenarios with moderate traffic, helping prevent outages by adjusting packet transmission historically.

% Upon receiving results from the \gls{dt}, the \gls{sadr} evaluates the test's reward function to determine if the configuration meets requirements or if \glspl{ue} data rates should be adjusted. The minimum reward safety threshold for \glspl{ue} is set to the average reward of scenarios with moderate traffic, aiding in outage prevention by adjusting packet transmission based on historical trials.

% fidelity of the system can be calculated as a loss function between the reward of the real system - the reward predicted from DT 
% Finally, when the control actions of the framework are run by both the DT and the real-world environment, we can measure the fidelity of the DT Twin and the loss of similarity of the system by calculating the Mean Squared Error (MSE) between the reward.
% The closer this value is to the zero, the closer the DT is to the real-world deployment.

\begin{algorithm}
\vspace{0.1cm}
\SetAlgoLined   
\SetKwProg{Upon}{upon}{:}{end}
\KwResult{Maximize $PSR_{i}(t)$ and guarantee the requested bandwidth of the UEs.}
$risk\_threshold \gets$ safety threshold triggering DT\;
$app\_requirements \gets$ minimum safety acceptable reward value\;
$safe\_setup \gets$ risk vector related to app\_requirements\;

\Upon{receive\_new\_ue\_traffic\_request(req)}{
$risk \gets compute\_risk(req.risk\_vector)$

\eIf{$risk > risk\_threshold$}{
       $twin\_evaluation(req)$\;
      }
      {
        $launch\_experiment(req.risk\_vector)$\;
      }
}

\Upon{twin\_evaluation\_completed(req, twin\_reward)}{
       \eIf{$twin\_reward >= app\_requirements$}{
        $reward = launch\_experiment(req.risk\_vector)$
        % $similarity\_loss = MSE(twin\_reward, reward)$
       }{
        $launch\_experiment(safe\_setup)$
    }
}  
 \caption{Safe Adaptive Data Rate main routine.}
 \label{alg:sadr}
\end{algorithm}
\vspace{-0.5cm}

\glslocalreset{dlai}
\subsection{\gls{dlai} Model Creation for Pilot Jamming}
Resource allocation stands out as a crucial aspect in wireless systems, yet resource constraints can sometimes pose challenges.
Certain systems may encounter limitations in allocating resources to specific areas due to their inherent capabilities. For instance, the dynamic creation or updating of \gls{dlai} models in real-time presents such a challenge. 
Imagine a scenario where a base station, tasked with managing channel pilot carriers, must stay alert to potential threats, such as pilot jamming.
% What is Pilot Jamming?
Pilot carriers in wireless networks aid in synchronization, channel estimation, and equalization. However, they're susceptible to interference, such as pilot jamming, which disrupts communication, leading to errors, reduced throughput, and compromised reliability. Combating pilot jamming requires robust signal processing and encryption techniques to protect pilot carriers.
Using a lightweight model for attack detection, offloading tasks such as \gls{dlai} model training to the cloud allows for dynamic resource allocation and continual improvement of spectrum sensing algorithms to accommodate changing network conditions, including possible pilot location shifts.

\begin{figure}[!t]
    \centering
    \includegraphics[width=.92\columnwidth]{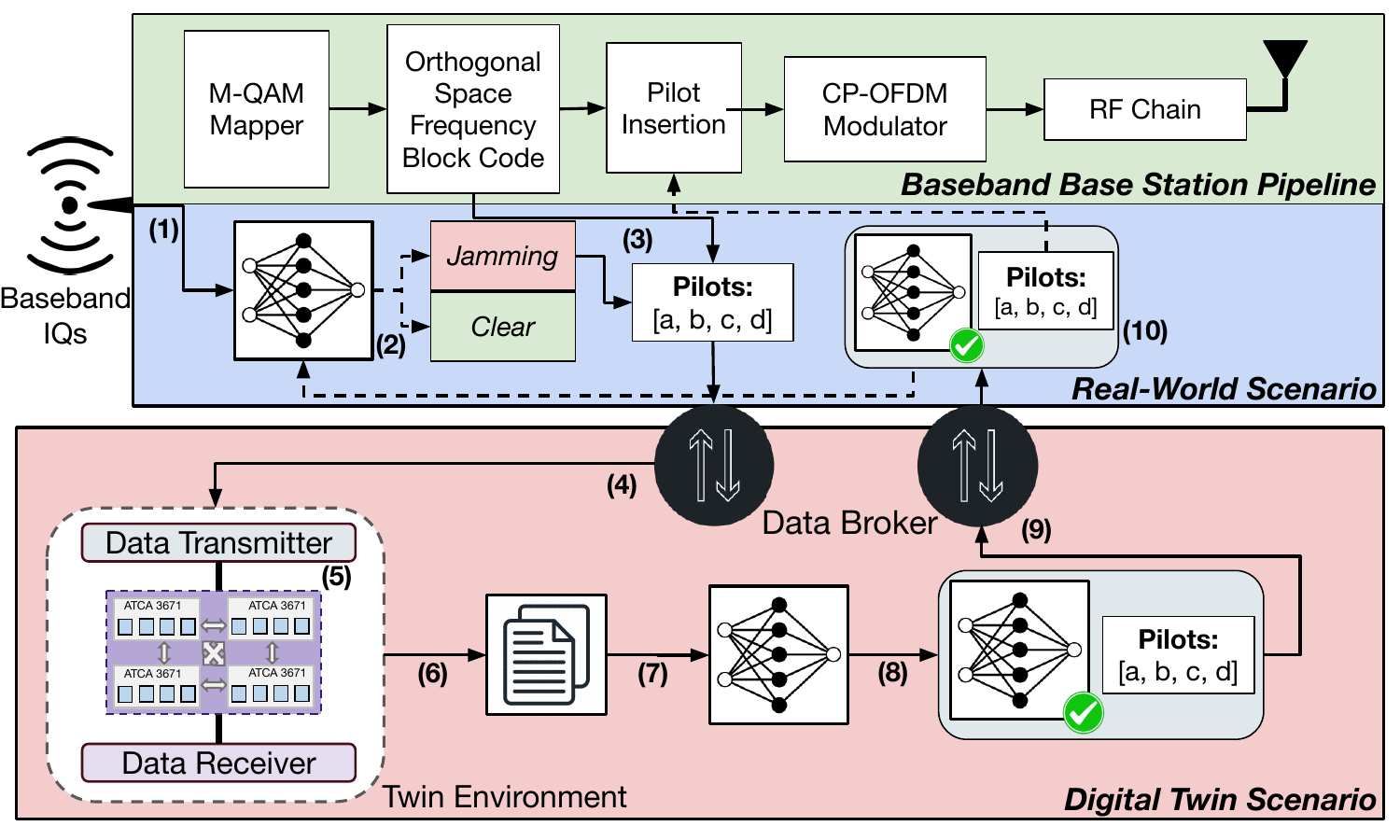}
    \vspace{-0.3cm}
    \caption{Overview of the pipeline where the base station requests a new jamming detection model based on the chosen pilot carriers.}
    \label{fig:pilot_pipeline}
    \vspace{-0.3cm}
\end{figure}

Fig.~\ref{fig:pilot_pipeline} provides a high-level overview of the pipeline involved when a base station requests a new \gls{cnn} model upon relocating its pilots. The process comprises three concurrent components: the base station pipeline, the physical layer jamming model, and the \gls{dt}. Firstly (1), the IQ samples are gathered and fed into the model for processing. Next, in (2), the \gls{cnn} output undergoes processing. If jamming is detected, (3) new pilot locations are chosen from the base station and forwarded to the \gls{dt} to request a new model. The \twinet broker transmits the new pilot information (4) to initiate the creation process, leveraging artificial data similar to that of a wireless jammer for training in (5). The data is then processed for model training (6), resulting in the generation of a new model (7). This new model, along with the updated pilots, is packaged and sent back to the base station (8) via the data broker. Upon model transfer, (10), the model is updated, and the new pilots are updated within the pipeline (9).

%%%%%%%%%%%%%%%%%%%%%%%%%%%%%%%%%%%%%%%%%%%%%%%%%%%%%%%%%%%%%%%%%%%%%%%%%%%%%%%%%%%%%%%%

\section{Experimental Results}
\label{sec:results}

In this section, we evaluate the performances of our adaptive control system and pilot jamming experiments. 
Sec.~\ref{subsec:sadr_res} discusses the results of the \gls{sadr} experiments, focusing on guaranteeing the best data rate possible for users.
Next, Sec.~\ref{subsec:dl} examines the pilot jamming experiments that were built from the pipeline in Fig.~\ref{fig:pilot_pipeline}, testing model creation time and generalized accuracy in over-the-air deployment.

\glslocalreset{sadr}
\subsection{\gls{sadr} for Traffic Management}
\label{subsec:sadr_res}

In all the experiments performed, all nodes are placed in static positions within the Arena testbed.
% Each \gls{ue} data rate request has been performed 
All the traffic from the radio point of view was transmitted in downlink on \gls{lte} Band 7 according to the \gls{3gpp} \gls{ts} standard 36.101~\cite{3gpp.36.101}.

Throughout the experiments, we test a specified range of traffic demand combinations for the \glspl{ue}, gradually increasing over time. Initially, the cumulative demanded data rate is minimal and considered safe, progressively escalating until the end of the experiment. As the experiment progresses, nearly all combinations exceed the application's security thresholds, requiring evaluation by \twinet. Each combination is tested for 60\,seconds, and every sequence of combinations, representing a unique experiment, is repeated 10\,times.

\begin{figure}[!t]
    \centering
    \includegraphics[width=.9\linewidth]{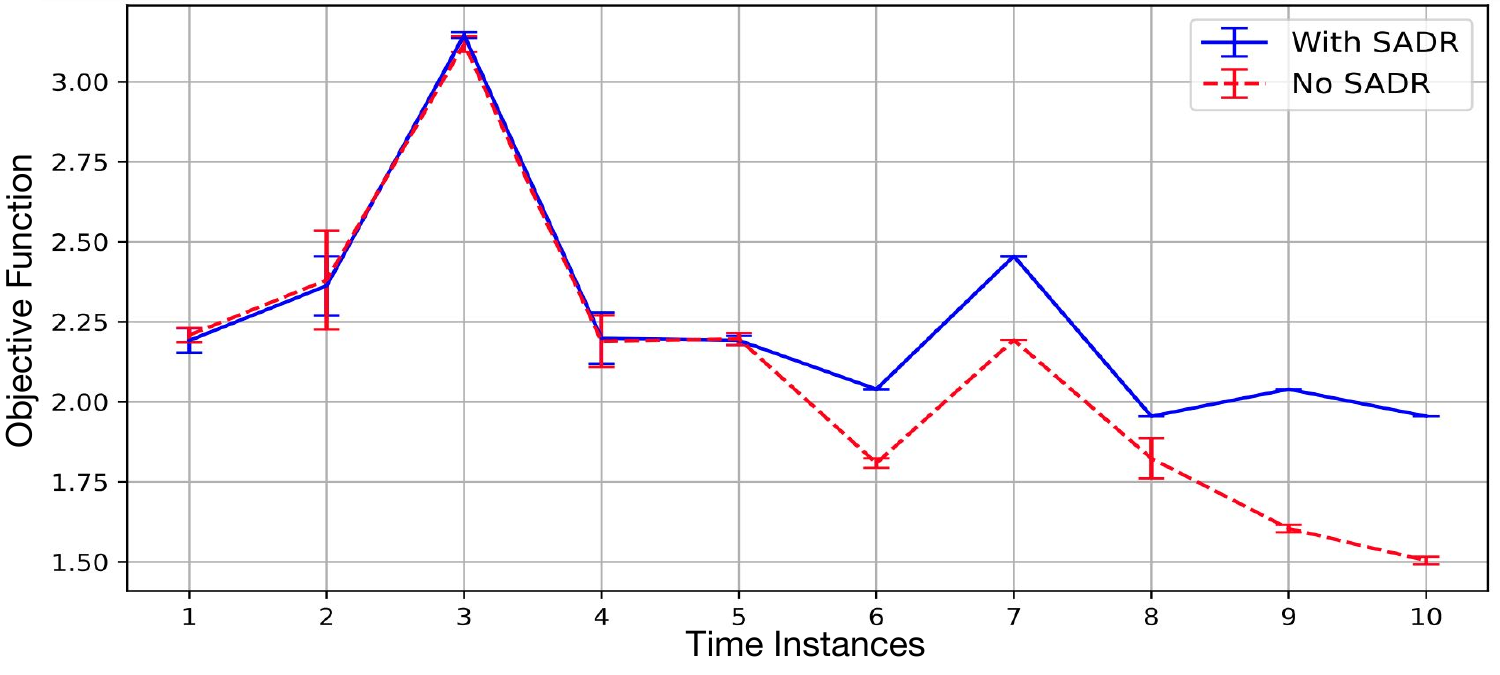}
    \vspace{-0.3cm}
    \caption{Comparison of the average reward function for experiments with and without \twinet. \Gls{sadr} data rate changes improve \glspl{ue} performance.}
    \label{fig:sadr_results}
    \vspace{-0.35cm}
\end{figure}

In Fig.~\ref{fig:sadr_results}, we depict the average objective function of experiments with and without risk assessment for user-requested traffic combinations. In one scenario, the \gls{sadr} system directly sends the possible risks to the \gls{dt}, which decides whether to alter the traffic combination.

In the plot analysis, we observe that as temporal instances progress, and thus, as user data transmission demands increase, the real system deteriorates without guidance. It attempts to forward all user traffic to the downlink to minimize the second component of the problem~(\ref{eqn:sadr_problem_formulation}). However, this exacerbates congestion in the radio channel, worsening both the \gls{psr} and the guaranteed traffic percentage for \glspl{ue}.
%From the analysis of the plot, we notice that as the temporal instances progress and, therefore, as the total amount of data transmission speed requested by the users increases, the real system, without any guidance or correction, tends to worsen its overall performance since it tries to send all the traffic requested by the users to the downlink, thus trying to minimize the second component of the objective function of the problem~(\ref{eqn:sadr_problem_formulation}), but due to the congestion created in the radio channel, this effort worsens both the \gls{psr} and the percentage of traffic guaranteed to the \glspl{ue}, i.e, the second component itself.
When assessed with the \gls{sadr}, the controller identifies high traffic versus the number of \glspl{ue}, which impacts guaranteed data rates per user. It dynamically boosts \gls{psr}, improving performance compared to the version without a \gls{dt}, highlighting the efficacy of the proposed algorithm and the potential of \gls{dt} with \twinet to optimize real network performance via virtual evaluation.
%When evaluated and adapted with the \gls{sadr}, the controller recognizes excessive traffic relative to the number of \glspl{ue}, compromising guaranteed data rate per user while dynamically increasing \gls{psr}, thus improving performance compared to the version lacking a \gls{dt}. This underscores the efficacy of the proposed algorithm and the potential of \gls{dt} with \twinet to enhance real network performance through prior evaluation in a virtual environment.

% \begin{figure}[!t]
%     \centering
%     \includegraphics[width=\linewidth]{figs/cm_twin_vs_real.pdf}
%     \caption{Confusion matrices showing the accuracy of the trained model on the \gls{dt} training validation data (LEFT) and the unseen, real-world data deployed in an over-the-air environment.}
%     \label{fig:cm_twin_real}
% \end{figure}

\begin{table}[!b]
\setlength\belowcaptionskip{5pt}
    \vspace{-0.45cm}
    \centering
    \footnotesize
    \setlength{\tabcolsep}{2pt}
    \caption{\gls{cnn} accuracy comparison by channel sizes indicating model performance on both \gls{dt} and Real-World datasets.}
    \label{tab:model_acc}
    \begin{tabularx}{\columnwidth}{
        >{\centering\arraybackslash\hsize=0.8\hsize}X 
        >{\centering\arraybackslash\hsize=0.8\hsize}X 
        >{\centering\arraybackslash\hsize=1\hsize}X
        >{\centering\arraybackslash\hsize=1\hsize}X }
        \toprule
        Channel Size (Bandwidth) & Pilot Amount & Training Accuracy (Twin) & Testing Accuracy (Real-World) \\
        \midrule
        10~MHz     & 4      & 0.999     & 0.981 \\
        20~MHz     & 4      & 0.998     & 0.964 \\
        40~MHz     & 6      & 0.963     & 0.941 \\
        \bottomrule
    \end{tabularx}
\end{table}

\glslocalreset{dlai}
\subsection{\gls{dlai} Model Creation for Pilot Jamming}
\label{subsec:dl}

We test a generalized approach to unique channel sizes by examining three cases for pilot jamming detection modeling seen in~\cite{robinson_narrowband_2023}, instead, using the VGG16 \gls{cnn}.
For our experiments, we utilize three subcarrier sizes: 10\,MHz for LTE networks, 20\,MHz for Wi-Fi, and 40\,MHz for Wi-Fi scalability testing.
Pilot carrier selection varies accordingly: for a 10\,MHz channel, 4\,pilots are chosen from 64\,subcarriers; for a 20\,MHz channel, 4\,pilots are selected from 128\,subcarriers; and for a 40\,MHz channel, 6\,pilots are selected from 128\,subcarriers.

% (156 KHz bandwidth), (156 KHz bandwidth), (312 KHz bandwidth)

% For a \gls{dlai} model, we selected the VGG16 \gls{cnn}. The \gls{cnn} architecture consists of 16 weight layers, including 13 convolutional layers followed by max-pooling layers and three fully connected layers, known for its deep architecture and simplicity in design. This model is useful in the wireless spectrum due to its deep architecture, which allows for hierarchical feature extraction, aiding in the identification of complex patterns and characteristics inherent in wireless signals. We converted the VGG16 to a 1-dimensional model with an input size of $(128,2)$, being the baseband IQ samples. Each model has an output size of the number of pilot subcarriers plus an extra output for normal spectrum activity. 

\begin{figure}[!t]
    \centering
    \includegraphics[width=.9\linewidth]{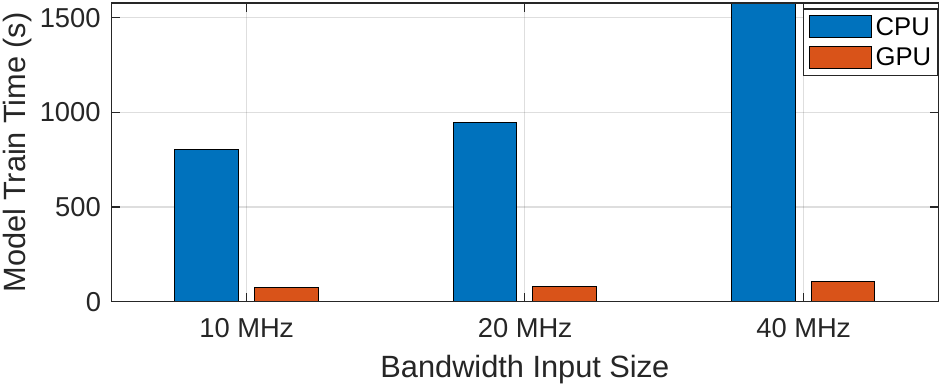}
    \vspace{-0.35cm}
    \caption{Time it takes to retrain new models based on channel bandwidth size on the fly within a \gls{dt} environment using both a CPU and GPU.}
    \label{fig:model_retrain_lat}
    \vspace{-0.35cm}
\end{figure}

Table~\ref{tab:model_acc} presents accuracy values for three models across different bandwidth sizes.
Training on artificial data from \gls{dt} yields consistently high accuracy, peaking at 99.9\% on the 10MHz channel and dropping to 96\% on the 40MHz channel. This aligns with expectations, as narrower bandwidths generally yield higher accuracy.
Post \gls{dt} training and validation, the \gls{cnn} model undergoes testing on over-the-air data, presenting a new challenge. Despite this shift, all models show only a modest 2.5\% average accuracy drop, indicating that \gls{dt} data closely mimics real-world conditions, facilitating reliable model deployment.
The model accurately identifies jamming locations in most data, though real-world testing may experience slight accuracy loss due to factors like low SINR or signal congestion.

Efficiently updating models with new pilots is crucial for adaptation and optimization. \gls{cnn} training is limited by computational resources and dataset size. We assess on-the-fly model generation on an Intel Xeon E5-2698 v4 @ 2.20GHz CPU and a NVIDIA Tesla V100 GPU. Training times for all models using both CPU and GPU are shown in Fig.~\ref{fig:model_retrain_lat}. GPU acceleration significantly reduces training times, with improvements of approximately 85.99\%, 86.90\%, and 90.75\% for Model 1, Model 2, and Model 3, respectively.

\begin{table}[!h]
\setlength\belowcaptionskip{5pt}
    \vspace{-0.6cm}
    \centering
    \footnotesize
    \setlength{\tabcolsep}{2pt}
    \caption{Time (in seconds) it takes to redeploy a newly trained model by channel size on both a CPU and GPU.}
    \label{tab:model_redeploy}
    \begin{tabularx}{\columnwidth}{
        >{\raggedright\arraybackslash\hsize=1.4\hsize}X 
        >{\raggedright\arraybackslash\hsize=0.8\hsize}X 
        >{\raggedright\arraybackslash\hsize=1\hsize}X
        >{\raggedright\arraybackslash\hsize=1\hsize}X
        >{\raggedright\arraybackslash\hsize=0.8\hsize}X
        >{\raggedright\arraybackslash\hsize=1\hsize}X }
        \toprule
        \multicolumn{6}{c}{\textit{\textbf{All times reported in seconds~(s)}}} \\
        \midrule
        Channel \newline Size & Data Transfer & Data Collection & Data Processing & Model Creation & Total Deployment \\
        \midrule
        10~MHz (CPU) & 0.039     & 42.030 & 2.418      & 803.404     & 847.891 \\
        10~MHz (GPU) & 0.039     & 42.030 & 2.418      & 74.643      & 119.130 \\
        20~MHz (CPU) & 0.041     & 47.209 & 2.627      & 946.312     & 996.189 \\
        20~MHz (GPU) & 0.041     & 47.209 & 2.627      & 81.175      & 131.052 \\
        40~MHz (CPU) & 0.044     & 42.093 & 2.659      & 1578.158    & 1623.954 \\
        40~MHz (GPU) & 0.044     & 42.093 & 2.659      & 105.486     & 150.282 \\
        \bottomrule
    \end{tabularx}
    \vspace{-0.3cm}
\end{table}

Table~\ref{tab:model_redeploy} shows the importance of GPUs in reducing redeployment time for efficient model deployment. Many base stations lack GPU functionality, causing delays in pilot jamming detection. Transitioning to a \gls{dt} environment with GPU support can cut deployment time by 87\%, minimizing downtime during jamming attacks.
% Table~\ref{tab:model_redeploy} illustrates the total redeployment time, highlighting the importance of GPUs for efficient model deployment. Currently, numerous base stations lack GPU functionality, leading to ineffective pilot jamming detection due to prolonged downtime awaiting model replacement. However, transitioning to a \gls{dt} environment with GPU support can reduce deployment time by an average of 87\%, significantly reducing potential downtime during jamming attacks.
%
The capability of \gls{dt}-emulated datasets in training models for diverse wireless spectra enhances redeployment effectiveness, significantly boosting threat detection against adaptable adversaries.

%%%%%%%%%%%%%%%%%%%%%%%%%%%%%%%%%%%%%%%%%%%%%%%%%%%%%%%%%%%%%%%%%%%%%%%%%%%%%%%%%%%%%%%%

\section{Conclusions}
\label{sec:conclusion}

In this paper, we introduced \twinet, a novel approach for synchronizing the real and digital worlds via the \gls{mqtt} protocol. Our framework enables scalable employment of \glspl{dt}, showcased in applications such as traffic monitoring, network management, and pilot jamming prevention. \twinet effectively mitigates cellular network outages by adjusting packet transmission while maintaining reasonable data rates. Future work will expand on resource-intensive implementations.

% Dont know how much future work we'll focus ob
% Future works on exploiting this approach for safe federated learning.

% For the \gls{sadr}, we acknowledge that one of the limitations of our framework is that its risk assessment capability is based on the a priori knowledge of a safety threshold that cannot be extracted without any historical knowledge of the experimental setup.
% Despite the use of \twinet and the \gls{dt} to derive an accurate estimate of this threshold, such value is proportional to the number of \glspl{ue}, so our future work will focus on the design and development of a regression function that can replace the risk threshold with a dynamic function that scales with the number of \gls{ue}, ensuring more flexibility and improving the \gls{sadr} risk assessment routine.
% Moreover, we plan to use the framework to perform parallel data collection to design an \gls{ai} agent that will be able to explore both risky actions and states that create outages on the virtual \glspl{ue}.
% Thanks to \twinet, the agent will base its decision on the observations vector composed by the \gls{kpm} coming from the lower layers of the cellular stack virtualized in the twin, potentially increasing the overall value of the reward function through the continuous control of the network while still aiming to minimize the \glspl{rlf} due to the overuse of the channel.

%%%%%%%%%%%%%%%%%%%%%%%%%%%%%%%%%%%%%%%%%%%%%%%%%%%%%%%%%%%%%%%%%%%%%%%%%%%%%%%%%%%%%%%%

\balance
\footnotesize
\bibliographystyle{IEEEtran}
\bibliography{bib}

\end{document}